\documentclass[preprint,showpacs,showkeys,superscriptaddress,nofootinbib]{revtex4-1} 
\usepackage[utf8]{inputenc}
\usepackage{cmap} 
\usepackage[T1]{fontenc}
\usepackage{microtype}
\usepackage{amsmath,amssymb,mathtools}
\usepackage{txfonts}

\usepackage{mathrsfs}
\usepackage{graphicx,grffile,textcomp}
\graphicspath{{.}}

\usepackage{textcomp}
\usepackage{booktabs,tabularx,dcolumn}
\newcolumntype{d}[1]{D{.}{.}{#1}}

\usepackage{url,hyperref}
\usepackage[usenames,dvipsnames]{xcolor}
\hypersetup{colorlinks=true, linkcolor=BrickRed, urlcolor=blue!50!black, citecolor=blue!50!black}

\usepackage[capitalize]{cleveref}
\crefrangelabelformat{equation}{(#3#1#4)$-$(#5#2#6)}

\usepackage{placeins}

\renewcommand\rho\varrho
\renewcommand\vec[1]{\textrm{\bfseries #1}}

\begin{document}
\title{Solvent coarsening around colloids driven by temperature gradients}

\author{Sutapa Roy}
\affiliation{Max-Planck-Institut f\"{u}r Intelligente Systeme,
   Heisenbergstr.\ 3,
   70569 Stuttgart,
   Germany}
\affiliation{IV. Institut f\"{u}r Theoretische Physik,
   Universit\"{a}t Stuttgart,
   Pfaffenwaldring 57,
   70569 Stuttgart,
   Germany}
 
\author{S. Dietrich}
\affiliation{Max-Planck-Institut f\"{u}r Intelligente Systeme,
  Heisenbergstr.\ 3,
  70569 Stuttgart,
  Germany}
\affiliation{IV. Institut f\"{u}r Theoretische Physik,
  Universit\"{a}t Stuttgart,
  Pfaffenwaldring 57,
  70569 Stuttgart,
  Germany}
   
\author{Anna Maciolek}
\affiliation{Institute of Physical Chemistry, Polish Academy of Sciences, 
Kasprzaka 44/52, PL-01-224 Warsaw, Poland}

\date{\today}
\begin{abstract}
Using mesoscopic numerical simulations and analytical theory we investigate 
the coarsening of the solvent structure around a colloidal particle emerging 
after a temperature quench of the colloid surface. Qualitative differences in 
the coarsening mechanisms are found, depending on the composition of the binary 
liquid mixture forming the solvent and on the adsorption preferences of the colloid. 
For an adsorptionwise neutral colloid, as function of time the phase being next 
to its surface alternates. This behavior sets in on the scale of the relaxation 
time of the solvent and is absent for colloids with strong adsorption preferences. 
A Janus colloid, with a small temperature difference between its two hemispheres, 
reveals an asymmetric structure formation and surface enrichment around it, even 
if the solvent is within its one-phase region and if the temperature of the colloid 
is above the critical demixing temperature $T_c$ of the solvent. Our phenomenological 
model turns out to capture recent experimental findings according to which, upon laser 
illumination of a Janus colloid and due to the ensuing temperature gradient between 
its two hemispheres, the surrounding binary liquid mixture develops a concentration 
gradient.
\end{abstract}
\maketitle

Coarsening is a paradigmatic example for non-equilibrium dynamics of systems approaching 
a steady state. Typically, this process is induced by a temperature quench of an initially 
homogeneous two-phase system, such as binary liquid mixture, polymer mixture, 
into the regime of immiscibility. The dynamics of coarsening, which has been studied intensively for the 
bulk fluid ~\cite{bray1994, binder2009}, is changed substantially by the presence of 
surfaces. In this latter context, a lot of attention has been paid to binary fluids 
in contact with planar surfaces in semi-infinite or film geometries. Strong 
efforts have been devoted to phase-separation guided by the surface, which occurs if 
-- as it is generically the case -- the surface prefers one species of the binary fluid 
over the other. Under such conditions, upon a temperature quench of an initially 
homogeneous system into the miscibility gap, plane composition waves propagate from the 
surface into the bulk and result in a transient layer structure \cite{binder2006}. There 
are numerous studies of this, so-called, surface-directed 
phase-separation process \cite{puri2005,tanaka2001}. Most of them assume that 
after a quench the system -- including its boundaries -- thermalize instantaneously so 
that the coarsening proceeds at constant temperature everywhere. Instead of quenching 
one can apply temperature gradients, e.g., by heating or cooling the boundary of a system. 
There is much less theoretical work concerning phase separation induced by temperature 
gradients, inspite of such conditions being created in various experiments and for 
practical applications, e.g., in polymer systems \cite{granick1996,cong1999,binder2013,
zhang2013, platten1995}. So far, the focus has been on systems bounded by ``planar'' and 
``neutral'' surfaces (i.e., with no preference for either component of a two-phase system), 
supplemented with boundary conditions which maintain a \textit{stationary} linear 
temperature gradient across a film \cite{gonnella2008,hong2010,binder2013, zhang2013}. The 
effects associated with a temperature quench of a boundary, whereupon the temperature gradient 
across the system varies in time, has rarely \cite{essery1990, araki2004} been considered, 
albeit for a planar geometry. To the best of our knowledge, in this context the effects due 
to spatio-temporal temperature gradients in fluids bounded by non-neutral surfaces, i.e., in the 
generic presence of surface fields, have not yet been explored.

Here we consider a spherical colloid suspended in a near-critical binary 
solvent, kept in its mixed phase above $T_c$. We study the dynamics of solvent coarsening 
following a temperature quench of the surface of the suitably coated colloid.
Our interest in this problem has been triggered by recent experiments with a partially 
gold-capped Janus colloid suspended in a mixed phase of water-lutidine mixture below its lower 
critical point \cite{bechinger2011, bechinger2016} in which, upon laser illumination with 
sufficient intensity, one observes phase separation of the solvent around the 
particle. The \textit{early} stage dynamics of this complex process has not yet been investigated.
A steady state  occurring at 
\textit{late} times has been considered in the studies of moving Janus colloids 
for quenches crossing the binodal \cite{wurger2015, samin2016}. In these studies the 
assumption has been made that the order parameter starts to evolve only after the stationary 
temperature profile has been reached. Here, we consider the \textit{simultaneous} time evolution of 
the coupled order parameter and temperature fields. This is expected to have repercussions for the 
motion of the Janus colloids\footnote{Allowing for the simultaneous evolution of coupled temperature 
and OP fields revealed that the body force exerted on a colloid due to the concentration flux 
is much stronger at the beginning of the coarsening process than in the stationary state. This 
suggests that the motion of the Janus particle may start before the stationary state is achieved.\label{fnote}}.
For homogeneous colloids we observe a surprising pattern evolution, which cannot be captured by 
the above assumption. For Janus particles we show that, unexpectedly, also temperature quenches, which 
{\bf do not} cross the binodal, lead to coarsening.

\begin{figure}
\centering
\includegraphics*[width=0.8\textwidth]{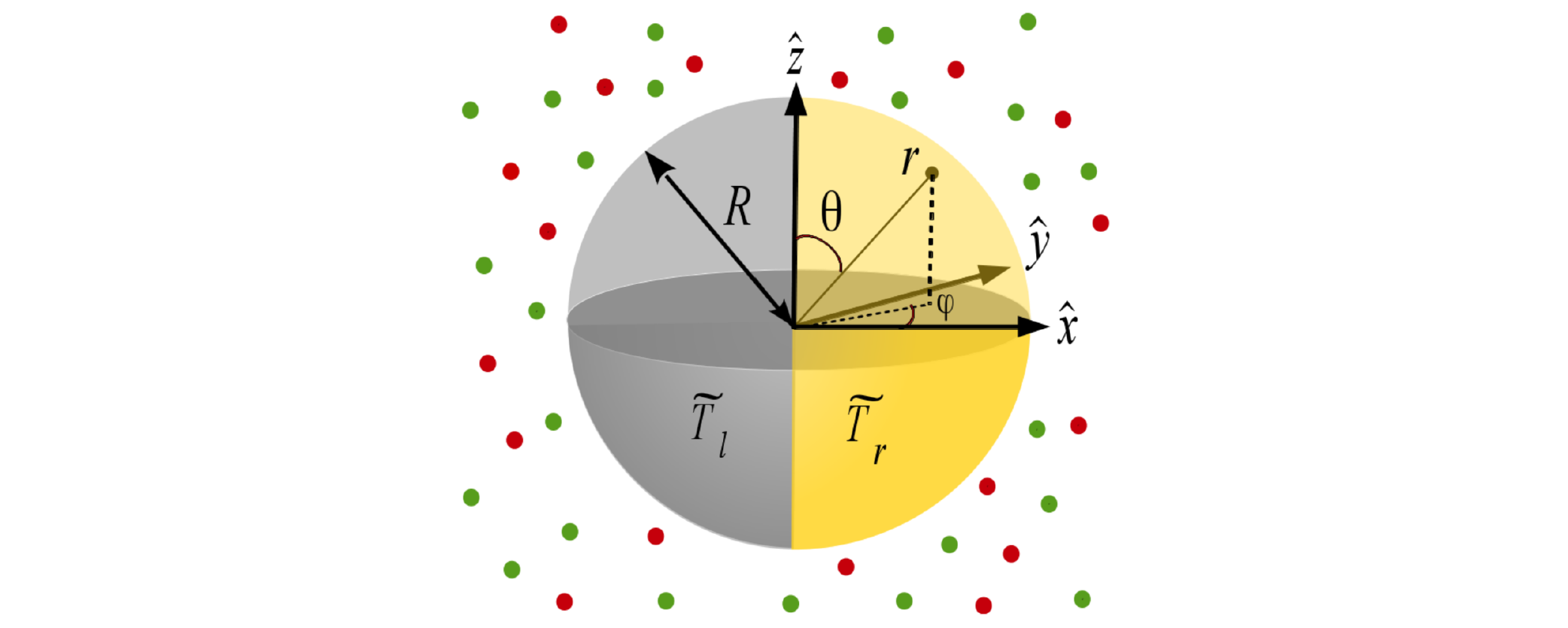}
\caption{Spherical Janus colloid of radius $R$ with reduced temperatures $\tilde T_r$ and 
$\tilde T_l$ on its right and left hemisphere, respectively. The azimuthal angle $\varphi$ 
is measured from the $x$-axis in the horizontal $\hat x$ $\hat y$ midplane of the colloid, 
the polar angle $\theta$ is measured from the $z$ axis, and $r$ is the radial distance from the 
centre. The initial temperature in the whole binary solvent is $\tilde T_i = \tilde T_r$.}
\label{fig1}
\end{figure}

We employ a phenomenological model, which we treat numerically and analytically. 
Specifically, we use the Cahn-Hilliard-Cook (CHC) type description, based on the Landau-Ginzburg 
free energy functional conjoined with the heat diffusion equation \cite{essery1990,bray1994}:
\begin{subequations}\label{CHC} 
\begin{align} 
&\frac{\partial \psi(\vec r,t)}{\partial t} = \nabla ^2 \Big ( \frac{{\tilde T}(\vec r,t)}{|{\tilde T}_1|} \psi(\vec r,t) + 
\psi^3(\vec r,t) - \nabla^2 \psi(\vec r,t) \Big)+\eta(\vec r,t),~~~\\
&\frac{\partial \tilde T(\vec r,t)}{\partial t} = {\mathcal D} \nabla ^2 \tilde T(\vec r,t).
\end{align}
\end{subequations}
Here $\psi(\vec r,t)$ is the local order parameter (OP) and $\tilde T(\vec r,t)$ is proportional to the 
reduced temperature field $(T(\vec r,t)-T_c)/T_c$. $T_1$ is the quench temperature of the colloid surface. 
The Gaussian random noise obeys the relation 
$\langle \eta(\vec r,t) ~\eta(\vec r', t')\rangle= -2\nu(\vec r) \nabla^2 \delta (\vec r-\vec r') \delta (t -t')$; 
$\nu(\vec r)$ is the strength of noise. Equation (\ref{CHC}) is valid for phase separation driven by 
diffusion, with hydrodynamic effects being irrelevant (e.g., for small P\'{e}clet numbers or at the early time of coarsening). 
${\mathcal D}={D_{th}}/(|\tilde T_1|D_m)$ involves the ratio of the thermal diffusivity $D_{th}$ of 
the solvent and the solvent interdiffusion constant $D_m$. \Cref{CHC} has to be complemented by boundary 
conditions (b.c.) on the surface ${\mathscr S}$ of the colloid. At a homogeneous 
surface the temperature field is constant, $\tilde T(\vec r)|_{\mathscr S} = {\tilde T}_1,$ and 
we assume there is no heat flux through the colloid. The generic preference of the colloid surface 
for one of the two components of the binary mixture is accounted for by the so-called Robin b.c. 
$({\hat n} \cdot \nabla \psi(\vec r) + \alpha \psi(\vec r))|_{{\mathscr S}}=h_s$ \cite{diehl1997}. 
Here, $\alpha$ and $h_s$ are the dimensionless surface enhancement parameter and symmetry breaking 
surface field, respectively. The second necessary b.c. \cite{diehl1992} is for no particle flux 
normal to the surface. \Cref{fig1} explains various notations. The spherical colloid of 
radius $R$ is placed at the center of a cubic simulation box (SB) with side length $L$ and periodic 
boundary conditions \cite{allen1987} at the side walls of SB. 
(For details concerning the model, numerical techniques, and relation to experimentally relevant quantities see SM). 
\begin{figure*}
\centering
\makebox[\textwidth]{\includegraphics[width=0.8\paperwidth]{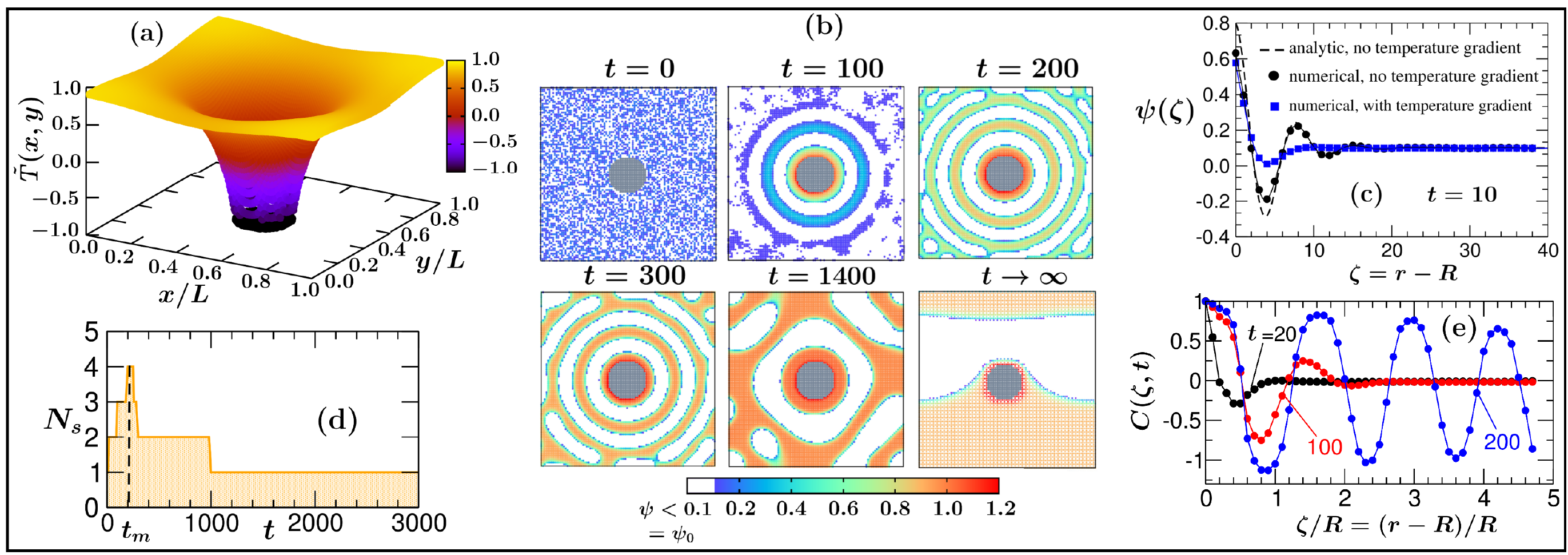}}
\caption{ Temperature-gradient induced demixing around a homogeneous colloid immersed in a binary solvent 
with off-critical concentration $\psi_0=0.1$ and undergoing slow cooling below $T_c$. 
The results in (a),(b),(d), and (e) correspond to $L=100,~R=10,
~\alpha=0.5,~h_s=1$, $\mathcal{D}=50$, and $\nu=10^{-4}$. (a) Temperature profile in 
the midplane $z=L/2$ at an early time $t=10$, exhibiting a strong gradient. Initially, the solvent 
is hot and the colloid is cold. 
(b) Coarsening patterns in the midplane of the colloid. Close to the colloid disconnected concentric 
circular structures form; away from the colloid spinodal-like patterns prevail. As time progresses 
the shells propagate into bulk via the formation of new layers. 
As expected, in the long time limit the system forms a planar interface trapping the colloid. 
(c) Comparison of the approximate analytic prediction of the OP profile without 
any temperature gradient ($(\bullet)$) and numerical data 
$(\color{blue}\blacksquare)$ with a temperature gradient, for $R=10$, $\tilde T_i=1$, $\alpha=0.01,~h_s=0.1$, 
and $\nu=10^{-4}$. 
(d) Non-monotonic time dependence of the number of concentric shells $N_s$; the decrease is much 
slower than the increase. 
(e) Radial two-point equal time correlation function $C(\zeta=r-R,t)$ vs. reduced distance $\zeta/R$ 
for three values of $t$. $C(\zeta,t)$ decays spatially fast at early times while with increasing time 
it develops multiple minima corresponding to concentric shell-like layers around the colloid. Data 
have been averaged over $10$ independent initial configurations.}
\label{fig2}
\end{figure*}

We first study demixing near a homogeneous colloid ($h_s=$ constant and $\tilde T_l=\tilde T_r=-1$) quenched from $\tilde T_i=1$ 
to $-1$. \Cref{fig2}(a) portrays a 
typical temperature profile in the midplane at an early time $t=10$. There is 
a strong temperature gradient, with $\tilde T(\boldsymbol r)$ near the side walls of SB being close 
to the initial value. As demonstrated in Figs. 2 and 3 in SM, this gradient reduces with time until 
the concentration profile attains a sinusoidal stationary state. 

In \cref{fig2}(b), we show a cross-sectional ($\hat x$ $\hat y$) view of the evolution patterns 
at six times. As the solvent cools a layered structure, consisting of concentric circular 
shells, forms near the colloid. Two neighboring layers contain opposite phases while the phase 
next to the colloid is $\psi > \psi_0$. Away from the colloid spinodal-like patterns prevail 
(see $t=100$). Upon increasing time, the shell structure propagates into bulk via formation of 
new layers and the maximal absolute value of the angularly averaged concentration in each layer 
increases (see Fig. 2 in SM).

Since a reliable analytic expression for OP in the presence of temperature gradients 
could not yet be obtained, numerical results are indispensable. In \cref{fig2}(c), we present 
numerical results ({\color{blue} $\blacksquare$}) for $\psi(r,t)$ after quenching a homogeneous 
colloid (not the solvent) to below $T_c$, i.e., with temperature gradient. The dashed line refers 
to our solution for linearized approximation of Eq.~(\ref{CHC}) without noise and 
temperature gradient, with the approximate form (by including $h_s$ in 
the calculations in \cite{essery1990,glotzer1999}): 
$\psi(\zeta=r-R,t)\approx \psi_0+\Big(-\alpha \psi_0+h_s)/(k_f^3 \sqrt{\pi t})\Big)\Big(R/(R+\zeta)\Big)
\exp\left(k_f^4 t-\zeta^2/(16k_f^2t)\right)\Big(A\cos~k_f\zeta-\left(\zeta/(4k_f^3t)\right)\sin~k_f\zeta \Big)$,
where $A=1+\left(5/(8 k_f^4 t)\right)\Big(1-\zeta^2/(8k_f^2t)\Big)$ with $k_f^2=(1-3\psi_0^2)/2$ 
characterizing the \textit{f}astest growing mode. Numerical data with temperature gradient 
({\color{blue} $\blacksquare$}) exhibit smaller peak compared to overall quench. The reason is: 
in case of a gradient temperature fronts propagate from colloid into bulk slowly and thus coarsening 
proceeds slowly. Thereby, at $t=10$, while the OP profile for an overall quench has already developed 
two prominent minima, for cooling it has acquired only one minimum the absolute value of which is 
also smaller.

Once the layers spread throughout the system they start 
to break up (see $t=1400$ in \cref{fig2}(b)). The non-monotonic behavior of the number of concentric 
shells $N_s(t)$ in \cref{fig2}(d) indicates a novel coarsening mechanism, likely due to an interplay 
of surface and bulk demixing. The growth and the decay of $N_s(t)$ are not symmetric about the time 
$t_m$ at which $N_s$ peaks; the break-up is slower.   

In order to further investigate the surface patterns around the colloid, we compute the radial 
two-point equal time correlation function in the midplane defined as $C(\zeta=r-R,t)=\langle \psi(R,t) 
\psi(R+\zeta,t)\rangle-\langle \psi(R,t) \rangle \langle \psi(R+\zeta,t) \rangle$. The symbol 
$\langle \cdot \rangle$ denotes the average over initial configurations of the angularly averaged 
$C(\zeta,t)$. For self-similar domains in bulk, $C(r,t)$ exhibits scaling \cite{roy2013}: 
$C_{\text{bulk}}(r,t) = {\mathscr C}(r/\ell(t))$, where ${\mathscr C}$ is for bulk and static 
equilibrium. $\ell(t)$ is mean domain size. In \cref{fig2}(e), $C(\zeta,t)$ is plotted vs. 
distance $\zeta/R$, for three $t$. While at early times $C(\zeta,t)$ decays spatially faster, upon 
increasing time spatial decay becomes less steep and $C(\zeta,t)$ develops multiple peaks corresponding 
to various surface layers. However, we could not find any data collapse for $C(r,t)$ onto a 
function of a single variable. This indicates \textit{non} self-similarity of coarsening patterns 
due to symmetry-breaking surface fields.
\begin{figure}
\centering
\includegraphics*[width=0.8\textwidth]{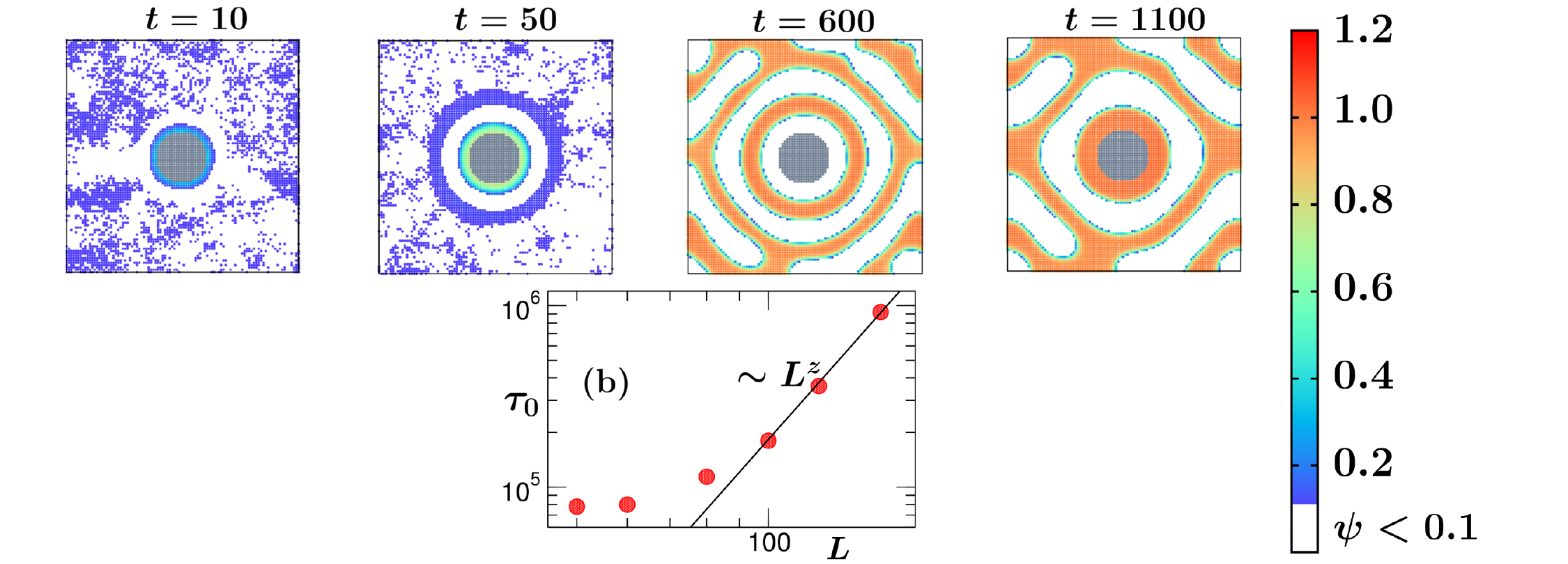}
\caption{(a) Snapshots of a cooling binary solvent $(\psi_0=0.1)$ around an adsorption-neutral, 
homogeneous colloid. At early times the phase $\psi > \psi_0$ is formed near the colloid 
surface until at $t\simeq 600$ the phase near the surface is replaced by the one with $\psi < \psi_0$. 
Results correspond to $L=100,~R=10,\tilde T_1=-1=\tilde T_l=\tilde T_r,~\alpha=0,~h_s=0,~\text{and} ~
\nu=10^{-4}$. (b) Plot of the crossover time $\tau_0$ vs. system size $L$ for $R=5$. Solid line drawn 
through the rightmost data points refers to the finite-size critical relaxation time 
$\propto L^z;~ z=4$ (see main text).}
\label{fig3}
\end{figure}

We have explored also the coarsening process around an adsorption-neutral colloid ($\alpha=0,~h_s=0$). 
For an off-critical solvent, the qualitative feature of the coarsening patterns for a neutral 
colloid (\cref{fig3}(a)) is similar to that with surface adsorption preferences (\cref{fig2}(b)). 
However, there is a difference concerning the phase formed at the colloid surface. While 
for $h_s > 0$ the phase with $\psi > \psi_0$ remains at the surface at all times, for the neutral 
colloid a crossover occurs: at very early times the phase $\psi > \psi_0$ is dominant near the 
surface until the layered structures have spread throughout the system and beyond a 
crossover time $\tau_0$ ($t=600$ in \cref{fig3}(a)) the phase $\psi < \psi_0=0.1$ is 
formed near the surface. We anticipate this first crossover time $\tau_0$ to be proportional to the OP 
relaxation time $\tau$ \cite{opbook} of the solvent. For a near-critical system $\tau \propto \xi^{^{~z}}$; 
$\xi$ being the equilibrium bulk correlation length. Thus for a finite system at $T_{c,\text{bulk}}$ 
it scales as $\tau \propto L^z$ \cite{folk2006}; $z \simeq 4$ is the dynamic critical exponent for 
model B \cite{folk2006} with diffusive dynamics for conserved order parameter (Eq. (1a)). 
In \cref{fig3}(b) data for $\tau_0$ are plotted for various system sizes. Agreement with the 
aforementioned power-law behavior (solid line) on a double-logarithmic scale supports the expectation 
$\tau_0 \sim \tau$. Such a crossover is observed only for neutral colloids, 
for off-critical concentrations, and for simultaneous time evolution of coupled OP and 
temperature fields. For the time evolution of the OP with stationary temperature profiles or for 
critical concentration, always both phases form near the surface and the OP morphology is not shell-like.
\begin{figure}
\centering
\includegraphics*[width=0.8\textwidth]{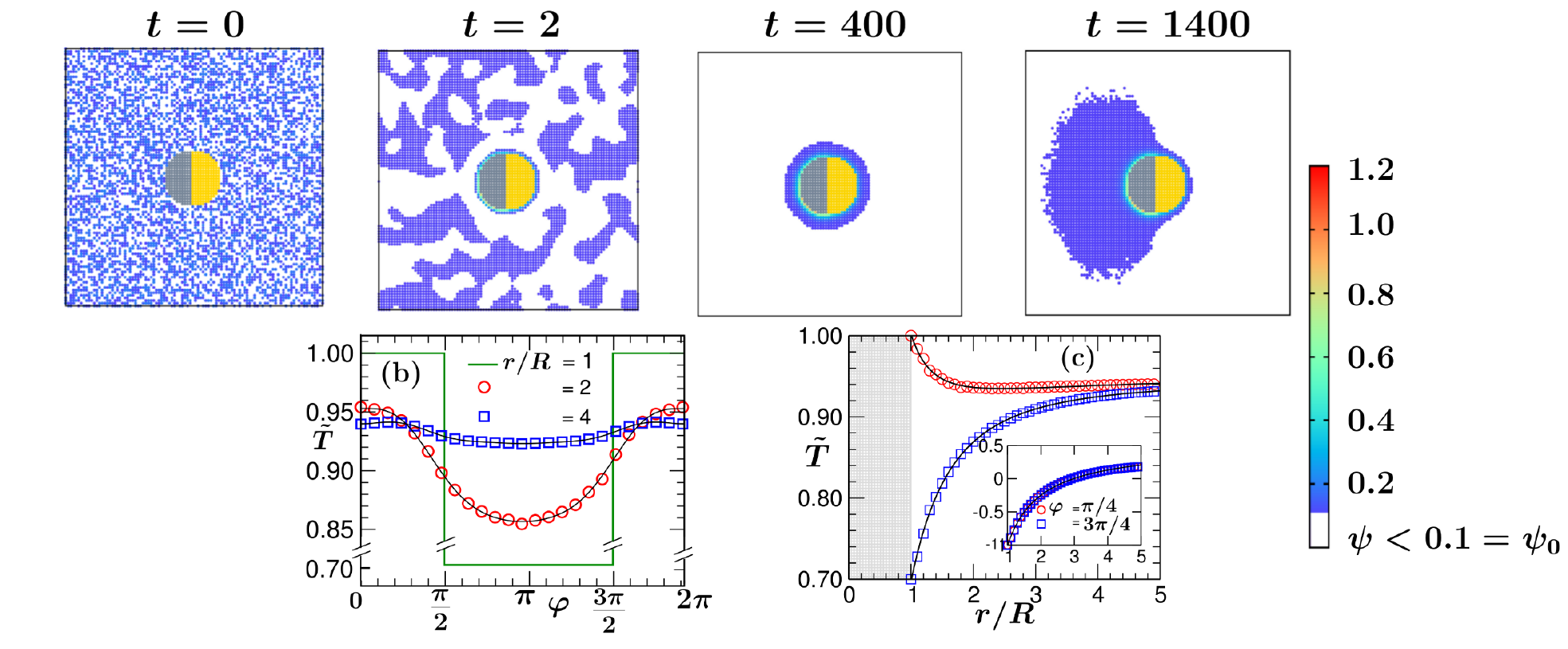}
\caption{Coarsening of a binary solvent with $\psi_0=0.1$ around a Janus colloid. Results 
correspond to $L=100,~R=10,~\alpha=0.5,~h_{s,l}=1,~h_{s,r}=0.5,~\text{and}~\nu=10^{-5}$. 
(a) Evolution patterns in its midplane ($\theta=\pi/2$). The left and right 
hemispheres are grey and yellow, respectively. Initially, both the solvent and colloid are at 
$\tilde T=1$. From $t=1$ to $400$ the system evolves at constant temperature everywhere. 
At $t=401$ the temperature of the left hemisphere is quenched to $\tilde T_l=0.7$.
The corresponding stationary configuration is shown 
at $t=1400$.
Although both hemispheres are maintained above $T_c$, structure formation is observed. 
(b) Temperature distribution $\tilde T(r,\varphi) > 0$ of the solvent in the midplane around a 
Janus colloid. Dependence (b) on the azimuthal angle $\varphi$ for two radial distances $r$ 
from the colloid centre and (c) on $r$ for two opposite angles $\varphi$ corresponding to the 
right and to the left hemisphere, respectively. The shaded region in (c) corresponds to the space 
occupied by the colloid. Inset of (c): as in main frame, but for the homogeneous colloid discussed 
in \cref{fig2} for which there is no dependence on $\varphi$.}
\label{fig4}
\end{figure}

Next, we turn to coarsening of solvent around a \textit{Janus} colloid with two 
hemispheres at different temperatures; both hemispheres 
prefer the same component of the solvent but with different strengths. \Cref{fig4}(a) 
shows OP distributions. We start with a homogeneous configuration of the binary solvent at 
$\tilde T=1$ and keep both hemispheres of the Janus colloid at $\tilde T=1$ (above $T_c$) at $t=0$. 
Subsequently, we let the system evolve ($t=1$ to $400$) such that there is no temperature gradient. 
Accordingly, the surface enrichment phenomenon is the only mechanism for structure 
formation. The snapshot at $t=400$ corresponds to the equilibrium surface adsorption OP profile. 
The Janus character of the colloid causes only weak deviations from a spherically symmetric adsorption 
profile. Next, at $t=401$ we quench the left hemisphere of the Janus colloid to $\tilde T=0.7$ such 
that the subsequent evolution will occur in the presence of a temperature gradient. The corresponding 
stationary configuration is shown at $t=1400$. The comparison of the snapshots at $t=400$ and $t=1400$ 
clearly demonstrates the difference between the equilibrium surface pattern formed due to surface 
enrichment only and the steady state pattern emerging in the presence of a temperature gradient. 
Clearly, a temperature gradient leads to a more pronounced bubble formation on the cold side of the 
Janus colloid. Coarsening in fluid regions with $T>T_c$ was observed experimentally \cite{granick1996} 
in polymer solutions due to the Soret effect and numerically \cite{araki2004} in fluid 
mixtures due to convective flows. Here, for {\it purely diffusive} dynamics, i.e., without involving 
any hydrodynamic flow, we observe the condensation of a droplet around the colloid above $T_c$, which is 
a novel phenomenon due to the combination of Soret and surface effects. 
To relate the anisotropy of surface patterns with temperature gradients, within 
the midplane we have computed the radial and angular dependence of $\tilde T$ on $r$ and $\varphi$, 
respectively. \Cref{fig4}(b) depicts the dependence of $\tilde T$ on $\varphi$ for two fixed 
values of $r$ with temperature in stationary state. The symbols correspond to our numerical 
data; the solid lines refer to analytical predictions \cite{wurger2013}: $\tilde T(r, \varphi)=
A_0+\sum_{n=0}^\infty B_n P_n(\cos \varphi) (R/r)^{n+1}$, where $B_n$ are constants and 
$\Big\{P_n\Big \}$ Legendre polynomials. In \cref{fig4}(c) we plot the radial dependence of $\tilde T$ 
for two opposite angles. Our numerical 
results agree with the theoretical predictions. The very slow (algebraic) decay of stationary 
temperature profile facilitates coarsening in extended regions of the system; in our simulation box 
it takes place everywhere. Due to finite size, away from the colloid $\tilde T$ is lower than its initial value.   
Note that $\tilde T(r,\varphi)$ is anisotropic, i.e., different for the two angles considered. This 
should be compared with the homogeneous colloid for which $\tilde T(r)$ is radially symmetric 
(see inset of \cref{fig4}(c)) and coarsening patterns are also symmetric (\cref{fig2}(b)). 
This confirms anisotropy of the temperature distribution to be the dominant source of the anisotropy 
in OP distribution around a Janus colloid. Upon increasing radius of the colloid, the radial extent 
of a stationary bubble of the phase preferred by the colloid and the amplitude of the OP profile increase moderately. 
The value of the OP at the left hemisphere surface is slightly larger than at a planar wall ($R\to \infty$). At 
the right hemisphere it is reduced to about half the value for planar wall. 
Interestingly, we find that the OP autocorrelation functions with 
time-dependent temperature gradients decay slower than in the case of the stationary temperature profiles.

In summary, the simultaneous time evolution of the coupled order parameter and temperature fields leads 
to new transient patterns. For deep quenches (corresponding to ${\cal D}\approx 100$), the 
time scale of patterns for a molecular solvent varies between 10$^{-1}$s and 10$^{-2}$s, depending 
on the size of the colloid and simulation box (see SM). There is anisotropic structure formation around 
a Janus colloid even if the colloid and the solvent are at temperatures corresponding to its one-phase 
region, which is different from {\it surface enrichment} \cite{Jones:89, puri2005, Frisch:93}. The former 
situation might have occurred in experiments in Ref.~\cite{bechinger2011, bechinger2016} for low intensity 
illuminations. Even if $\tilde T(\mathbf{r},t)$ exhibits a fast dynamics, the results of our study are 
relevant for controlling pattern formation, e.g., in polymers, using hot homogenous or Janus particles 
\cite{Kurita:2017}. It is expected that the transient dynamics crucially influences the final patterns; 
these patterns should definitely be different from those seen for spatially homogeneous quenches. 
Our study contributes to the understanding of the propulsion mechanism via diffusive dynamics for one of 
the commonly used representatives of synthetic active matter \cite{bechinger2011,bechinger2016} (see (\ref{fnote})). 
The other presently available theoretical approaches \cite{wurger2015,samin2016} are based instead on 
hydrodynamics and are complementary to our study. Generally, dipping a particle into its binary solvent 
causes transportation of the preferred phase towards its surface. From our study we conclude that this 
transport is strongly enhanced if supported by a time-dependent temperature  gradient.  

\textbf{Acknowledgments:} The work by AM has been supported by the Polish National Science Center 
(Harmonia Grant No. 2015/18/M/ST3/00403).

\end{document}